\documentclass[letterpaper, 10 pt, conference]{ieeeconf}  

\IEEEoverridecommandlockouts                              
\usepackage{times}
\usepackage{helvet}
\usepackage{courier}

\usepackage{color}
\usepackage{amsmath}
\usepackage{amssymb}
\usepackage[ruled,vlined,linesnumbered]{algorithm2e}
\usepackage{graphicx}
\usepackage{epsfig}
\newtheorem{defi}{Definition}

\newtheorem{lemma}{Lemma}
\SetKwFor{Loop}{Loop}{}{EndLoop}

\title{\LARGE \bf Correlation Clustering Based Coalition Formation\\ For Multi-Robot Task Allocation}
\author{Ayan Dutta$^1$, Vladimir Ufimtsev$^2$, Asai Asaithambi$^1$
\thanks{$^1$School of Computing at the University of North Florida, USA.
{\tt\small \{a.dutta, asai.asaithambi\}@unf.edu}}%
\thanks{$^2$Dept. of Mathematics \& Computer Science at the East Central University, USA.
{\tt\small vufimtsev@ecok.edu}}
}
\long\def\omitit#1{}

\begin{document}
\maketitle
\thispagestyle{empty}
\pagestyle{empty}

\begin{abstract}
In this paper, we study the multi-robot task allocation problem where a group of robots needs to be allocated to a set of tasks so that the tasks can be finished optimally. One task may need more than one robot to finish it. Therefore the robots need to form coalitions to complete these tasks. Multi-robot coalition formation for task allocation is a well-known NP-hard problem. To solve this problem, we use a linear-programming based graph partitioning approach along with a region growing strategy which allocates (near) optimal robot coalitions to tasks in a negligible amount of time. Our proposed algorithm is fast (only taking 230 secs. for 100 robots and 10 tasks) and it also finds a near-optimal solution (up to 97.66\% of the optimal). We have empirically demonstrated that the proposed approach in this paper always finds a solution which is closer (up to 9.1 times) to the optimal solution than a theoretical worst-case bound proved in an earlier work. 
\end{abstract}

\section{Introduction}
Due to the inherent complexity of real world tasks (e.g., search and rescue, fire extinguishing, information collection) and limited capabilities of the available robots in the market, it is almost impossible for one robot to finish a complex task. As a result, cooperation among the robots is one of the basic requirements for any task completion. One form of cooperation among robots is coalition formation, where a subset of available robots forms a team that is assigned to a specific task. 
In this paper, we do not study how the task is subdivided among the members of a robot coalition. Instead, we study how a set of $N$ robots can be optimally divided into $M$ coalitions to complete $M$ tasks. Most of the previous research in coalition formation deals with software agents \cite{michalak2010distributed,Rahwan08,ramchurn10}. But 
real-world complexities, such as the on-board computational power of the robots and constrained communication, tend to limit the number of robots these algorithms can handle. The approach of this paper successfully handles as many as 100 robots. 

The problem of coalition formation for multi-robot task allocation can be described as follows: given a set of $M$ tasks and $N$ robots ($N > M$), how to partition the robot group optimally into coalitions where each coalition will be assigned to a unique task. It has been proven that multi-robot coalition formation problem for task allocation is NP-hard to both solve exactly and to approximate within a factor of $O(M^{1-\epsilon}), \forall \epsilon>0$ \cite{adams2011coalition}. Therefore, solving the coalition formation problem for multi-robot task-allocation in a reasonable amount of time while retaining the quality of the formed coalitions is an extremely challenging problem. In this paper, we propose an efficient algorithm for coalition formation by a group of mobile robots for task allocation so that the algorithm can run on the robots' on-board computers. 
We present a novel {\em value} function for the tasks which is used to assign only the required number of robots to each task. We also present a distance-based {\em cost} function to minimize the travel distances by the robots to the tasks. Our approach employs a clustering-based coalition formation methodology \cite{bansal2004correlation}. Clustering is a technique of gathering `common' elements under the same label. We exploit this idea to allocate nearby robots considered as `common' to the same cluster, centered around a specific task, while distant robots are assigned to different tasks. Results show that our approach finds a near-optimal solution (up to $97.66\%$ of the optimal) in a negligible amount of time. 

\section{Related Work}
Autonomous robots need to cooperate with each other to complete complex tasks at hand. Forming teams (or, coalitions) for efficient task completion is a computationally hard problem. One of the earliest studies on task allocation via coalition formation is due to Shehory and Kraus \cite{Shehory98}, in which the authors have proposed a greedy algorithm that is guaranteed to find a solution within a factor of $(k + 1)$ of the optimal where $k$ is maximum size of any coalition formed. Coalition formation by a multi-agent system has been studied extensively in the following decade. Optimal \cite{Rahwan09, rahwan2007anytime} and near-optimal \cite{rahwan2007near} solutions for coalition formation have been proposed. Most of these proposed algorithms employ a search-based technique to find the best solution. Although coalition formation algorithms have been developed frequently in the last decade, very few of them are targeted for multi-robot/agent task allocation \cite{adams2011coalition}. Taxonomies of coalition formation algorithms for task allocation  are proposed in \cite{lau2003task, gerkey2004formal}. Following the taxonomy in \cite{gerkey2004formal}, our work in this paper can be classified as addressing a single-task robot and multi-robot task problem. In other words, each robot is capable of completing only one task at a time and each task requires more than a single robot to finish. A distributed solution which formulates the coalition formation problem for multi-agent task allocation as a distributed set partitioning problem is proposed in \cite{tosic2004maximal}. In \cite{adams2011coalition}, the authors have proposed a modified version of the algorithm proposed in \cite{Shehory98} and the complexity of their algorithm, ($O(N^{\frac{3}{2}}M)$), is polynomial compared to that of Shehory and Kraus \cite{Shehory98}, which is $O(N^kM)$, exponential on the size of the largest coalition. However, both \cite{adams2011coalition,Shehory98} report similar sub-optimality guarantees. 

Our proposed solution in this paper generates coalitions using a correlation clustering technique \cite{bansal2004correlation,demaine2003correlation}. It is a very commonly used technique in machine learning and pattern recognition. In correlation clustering, highly correlated points, robots in our case, are assigned to the same clusters whereas the points with low correlation are allocated to different clusters. One cluster is formed centering on one specific task and the result of this clustering process is equivalent to the generation of non-overlapping coalitions of robots.


\section{Problem Definition and Notations}
Let $R = \{r_1,r_2,..,r_N\}$ denote a set of $N$ robots. Each robot is characterized by a tuple $\langle P_i,\theta_i \rangle$ where $P_i$ and $\theta_i$ respectively denote the position and the orientation of the robot $r_i$. We assume that each robot is able to localize itself in the environment using an on-board GPS. Let $T = \{t_1,t_2,..,t_M\}$ denote a set of $M$ tasks ($N > M$). Any task $t_j$ is characterized by a tuple $\langle P_j,O_j \rangle$ where $P_j$ and $O_j$ respectively denote the task location and optimal number of robots needed to finish that task. The value of $O_j$ for each task $t_j$ is pre-defined and this information is assumed to be available with the robots. We assume that $\sum_{{1 \leq j \leq M}} O_j = N$. The robots are homogeneous in nature, i.e., any robot can be exchanged with any other robot. The environment is assumed to be a rectangle of size $length \times width$ is discretized into a set of square cells (denoted by $cell$) and one cell can only be occupied by at most one robot or one task at a time.

A {\em coalition} $c \subseteq R$ is a team of robots assigned to one task. 
Without loss of generality, we sometimes call a coalition a cluster. A {\em coalition structure} $CS$, defined as a partition, can be thought of as a set of non-overlapping clusters which covers all the robots. 
Let $CS=\{c_1,c_2,\cdots,c_M\}$ denote a coalition structure with $c_i$ assigned to task $t_i$ for $i=1,2,\cdots,M$. Let $\zeta$ denote the set of all possible partitions, and hence $CS \in \zeta$.
\newline
\textbf{Value Function} As the robots are homogeneous, the effectiveness of any robot coalition depends solely on the size of the coalition. We define $Val: CS \rightarrow \mathbb{R}$, a {\em value function} that 
assigns a virtual reward to a coalition and is defined as
\begin{equation}
Val(c_i) = O_i^2 - (O_i - |c_i|)^2.
\label{eqn_task_val}
\end{equation}
$Val$ ensures that if the coalition $c_i$ assigned to a certain task, $t_i$, has $O_i$ members in it, i.e., $|c_i| = O_i$, then that coalition earns the maximum possible value. If the size of the coalition is greater or less than the associated $O_i$, then the value of the coalition is not the highest. On the other hand, if $|c_i| > 2|O_i|$, then the value of the coalition becomes negative. This makes sure that none of the formed coalitions is too large in size if it is not required by the pre-defined $O$ value. We define the value of a coalition structure as the summation of values of all the coalitions in it, i.e., $Val(CS) = \sum_{\forall c_i \in CS} Val(c_i)$. Note that the maximum value of any coalition structure can be mathematically computed as follows: $MAX\_VAL = \sum_{{1 \leq j \leq M}} O_j^2$.\newline
\textbf{Cost Function }The robots are initially randomly placed in the environment. When a robot is assigned to a task as part of a coalition, it needs to move to the task location to complete the task. Each robot spends a certain amount of energy (in terms of battery power) to move from one point to another. This is represented using the proposed {\em cost-distance} function, defined as $cost_{dist}(r_i, t_j) = \frac{d(P_i, P_j)}{\sqrt{length^2+width^2+1}}$ where $d$ denotes the Euclidean distance between two locations. 
We next define the quantity $f_{val}(r_i, t_j)=1-cost_{dist}(r_i, t_j)$ to represent the probability 
of a pair of robots or robot-task pair being in the same coalition. From here, we can calculate the `similarity' between a task and a robot \cite{demaine2003correlation} as 
\begin{equation}
w(r_i, t_j) = \ln\left(\frac{f_{val}(r_i, t_j)}{1-f_{val}(r_i, t_j)}\right).
\label{eq_edge_weight}
\end{equation}
We use the same function to represent the similarity between two robots $r_i, r_j$. A higher value of $w$ indicates that the members of the pair of robots or the robot-task pair are `similar' and they should be in the same coalition, while a lower value of $w$ would mean that they are `dissimilar' and should be in different coalitions. To ensure that no two tasks are part of the same coalition, we define their similarity to be highly negative.
\newline
\textbf{Problem Objective} The problem objective is to find a set of coalitions for all the tasks such that the generated coalition structure has the minimum cost, while its value is the maximum. For each coalition $c_i \in CS$ assigned to task $t_i$, a \textit{cohesion} function is defined as follows: 
\[Co(c_i) = \sum\limits_{r_j \in c_i}w(r_j, t_i) + \sum\limits_{r_j,r_l \in c_i, j\neq l}w(r_j, r_l)\] and the cohesion quality of $CS$ is
\[ CQ(CS) = \sum\limits_{\forall c_i \in CS} Co(c_i).\]
Now we can formally define the multi-robot task allocation problem as follows:
\begin{defi}
Given a set of $N$ robots and $M$ tasks and each task $t_i$ requiring $O_i$ number of robots to finish it, find the coalition structure $CS^*$ containing $M$ coalitions (to be assigned to $M$ tasks) where:
\begin{align*}
  CS^* &= \arg \max_{CS \in \zeta} CQ(CS),\\
  \noalign{also satisfying}
  Val(CS^*) &= MAX\_VAL.
\end{align*}
\end{defi}

\section{Coalition Formation Algorithm\\ for Task Allocation}

The total number of possible coalition structures (partitions) is exponential in the number of robots. For $N$ robots, and a fixed size $M, 1\leq M \leq N$, the total number of coalition structures containing exactly $M$ non-empty coalitions is given by the Stirling number of the second kind: $S(N,M) = \frac{1}{M!}\sum_{i=0}^{M}(-1)^{i}\binom{M}{i}(M-i)^N$. 
Thus the number of possible coalition structures grows exponentially in the number of robots. With the goal of reducing the complexity of finding the optimal coalition structure, we use the framework of \cite{demaine2003correlation} which models the set of robots and tasks as a weighted complete graph. The robots and tasks are represented by vertices of the graph and edge weights correspond to the `similarity' of a pair of robots or robot-task being in the same coalition. The cohesion quality of a given coalition structure (partition of robots into coalitions) is calculated by summing the edge weights of all edges that are between robots in the same coalition. If two robots are in different coalitions, the weight of the edge between them is not included in the sum. To actually generate a coalition structure, we use a graph partitioning algorithm to split the vertices (robots) into groups (coalitions) under the constraint that the generated coalition structure has close to optimal cohesion quality. 

\subsection{Linear Programming Formulation for Graph Partitioning}

For our purposes, $G = (A, E, w)$ will be an undirected, weighted, complete (fully-connected) graph. $A$ is the set of vertices which corresponds to the set of robots $R$ and set of tasks $T$ i.e. $A=T\cup R$, and $E$ is the edge set which consists of all possible pairs of robots and tasks from $A$ (thus $|E| = \binom{|R|+|T|}{2}$). The edge weight function $w:E\rightarrow \mathbb{R}$ is as defined in Eq. (\ref{eq_edge_weight}). 





For any given coalition structure $CS$, the \emph{penalty} is defined by
\begin{equation}
Pen(CS) = Pen_p(CS) + Pen_m(CS),
\end{equation}
where $Pen_p(CS)$ corresponds the sum of positive edge weights across different coalitions and $Pen_m(CS)$ corresponds to the sum of negative edge weights within the same coalitions. More specifically:
\begin{equation}
Pen_p(CS) = \sum\limits_{\substack{w(r_i,r_l)>0 \\ r_i\in c_{k_1},r_l\in c_{k_2},\\ k_1\neq k_2}}|w(r_i,r_l)| +\sum\limits_{\substack{w(r_i,t_j)>0 \\ r_i\in c_{k_1} \\ k_1\neq j}}|w(r_i,t_j)|
\end{equation}
\vspace{-0.1in}
\begin{equation}
Pen_m(CS) = \sum\limits_{\substack{w(r_i,r_l)<0 \\ r_i,r_l\in c_{j}}}|w(r_i,r_l)| + \sum\limits_{\substack{w(r_i,t_j)<0 \\ r_i\in c_{j}}}|w(r_i,t_j)| 
\end{equation}

Note that the subscript of a coalition matches the subscript of the task it is assigned to i.e. coalition $c_j$ corresponds to task $t_j$. The penalty incorporates both positive weighted edges between \emph{different} coalitions and negative weighted edges that are part of the \emph{same} coalition. Through the maximization of the sum of edge weights within coalitions, the optimal coalition structure is obtained, considering only the function $CQ(CS)$, without the $Val(CS)$ function. Through minimization of the penalty, the cohesion quality function ($CQ(CS)$) of the coalition structure is maximized \cite{dasgupta2012dynamic}.

As specified in \cite{demaine2003correlation}, for each edge $e=(a_i,a_j)\in E$, where $a_i,a_j \in A$, binary variables $x_{a_i,a_j} \in \{0,1\}$ for a clustering (coalition structure) $CS$ are defined as: $x_{a_i,a_j}=0 \leftrightarrow \exists c_l \in CS: a_i,a_j\in c_l$ (i.e., $a_i,a_j$ are in the same coalition) and $x_{a_i,a_j}=1 \leftrightarrow \exists c_{k_1}, c_{k_2}\in CS, k_1\neq k_2: a_i\in c_{k_1}, a_j\in c_{k_2}$ (i.e., $a_i,a_j$ are in different coalitions). We will use $x_{a_i,a_j}$ and $x_e$ interchangeably from here on. The $Pen(CS)$ is then reformulated using the following non-negative constants:
$$
m_e= \left\{
\begin{array}{cc}
|w(e)|  & \text{if }w(e) < 0 \\
0 & \text{if }w(e) \geq 0
\end{array}
\right.
$$

$$
p_e= \left\{
\begin{array}{cc}
|w(e)|  & \text{if }w(e) > 0 \\
0 & \text{if }w(e) \leq 0
\end{array}
\right.
$$

$Pen(CS)$ is then given as:
\begin{equation}
Pen(CS) = \sum\limits_{e\in E}p_e x_e + \sum\limits_{e\in E}m_e(1-x_e)
\label{eqn_penalty}
\end{equation}

As stated previously, finding a coalition structure with minimal penalty is equivalent to finding the structure with maximal cohesion quality $CQ(CS)$. This problem is given as the following $0$-$1$ integer linear program:\\

min: $\sum\limits_{e\in E}p_e x_e + \sum\limits_{e\in E}m_e(1-x_e)$\\

constraints:\\ 
$x_{a_i,a_j} \in \{0,1\}, \forall a_i,a_j\in A, i\neq j$\\
$x_{a_i,a_j}+x_{a_j,a_k} \geq x_{a_i,a_k}, \forall a_i,a_j,a_k \in A, i\neq j \neq k$\\
$x_{a_i,a_j}=x_{a_j,a_i} \forall a_i,a_j\in A, i\neq j$\\

The second constraint is the triangle inequality, while the third is the symmetry constraint. These ensure that a valid coalition structure is generated from the solution. Since this problem is NP-complete, it is relaxed to a linear program with the same objective function and the following constraints: ~\cite{demaine2003correlation,dasgupta2012dynamic}:\\ 

$x_{a_i,a_j} \in [0,1], \forall a_i,a_j\in A, i\neq j$\\
$x_{a_i,a_j}+x_{a_j,a_k} \geq x_{a_i,a_k}, \forall a_i,a_j,a_k \in A, i\neq j \neq k$\\
$x_{a_i,a_j}=x_{a_j,a_i} \forall a_i,a_j\in A, i\neq j$\\

Algorithm \ref{linear prog} shows the pseudo-code for the coalition structure formation based  only on the cohesion quality. This process runs in polynomial time (in $N$) and gives a $O(\log N)$ approximation (see \cite{demaine2003correlation}). Although this problem can be solved in polynomial time, the solution may be non-integer i.e. fractional. Note that, if $0 < x_{a_i,a_j} < 1$, then there is no definite answer and we can think of $x_{a_i,a_j}$ as the probability that $a_i,a_j$ are in different coalitions. In this case, there is extra work to be done in order to determine whether or not $a_i,a_j$ are in the same coalition. This `rounding off' procedure is explained in the next section (IV.B Region Growing). It may also happen that some robots are not assigned to any coalition where there is a task. They may be in their own singleton coalition, or may be in a cluster with other robots but no task. In this case, extra work also has to be done to assign the robots to the best possible coalition. In fact, in such a case, there will be coalitions with tasks that will not have a sufficient amount of robots to complete the task, since in our formulation it is assumed that the total number of robots needed to complete all tasks is exactly $N$. The region growing technique explained in the next section can also be used in these scenarios. 
Another situation to consider is that even if an integer solution is obtained, the value of the coalition structure found may not be the maximum ($MAX\_VAL$), meaning some coalitions will have too many or too few robots. In this case also, we use the region growing algorithm to optimize the value.

\begin{algorithm}
\small{
\KwIn{$R$: A set of robots;\\ $T$: A set of tasks.}
\KwOut{$CS$: A coalition structure;\\ $\mathcal{R}_{ua}$: A set of unassigned robots.}
$\mathcal{R}_{ua} \leftarrow \emptyset$\\ 
\For{each $(a_i, a_j) \in A, (A = T \cup R)$}
{
Calculate $w(a_i,a_j)$.
}
Set the linear program constraints after calculating the penalty function (Eq. \ref{eqn_penalty})\\
Obtain a solution that satisfies the above-mentioned constraints by solving the linear program.\\ 
\eIf {$0$-$1$ integer solution is obtained}
{
Whenever $x_{a_i,a_j}=0$, group $a_i,a_j$ into the same coalition. This will create a valid coalition structure $CS$ (due to the symmetric and triangle inequality constraints).
\\
\eIf{$Val(CS) \neq MAX\_VAL$}{Use the Region Growing algorithm (Algorithm \ref{region growing})}{return $CS$.}
}
{Add the robots, for which all the edges including them yields a fractional solution, to $\mathcal{R}_{ua}$.\\
Use the Region Growing algorithm (Algorithm \ref{region growing}).}
    \caption{Coalition structure formation based on the $CQ$ function}
    \label{linear prog}
    }
\end{algorithm}



\subsection{Region Growing}
The coalition structure found by the Algorithm \ref{linear prog} maximizes the cohesion quality of $CS$, but does not take the value of $CS$ into consideration.
For this reason, it might so happen that one of the coalitions formed in this stage is unnecessarily large and as a consequence, while other coalitions may be smaller in size than required. 
For instance, let us suppose that in a warehouse, $M$ stacks of boxes need to be moved from one place to the other. In this case, each stack of boxes needs four robots to carry it, because otherwise, either the stack will fall or it is probably too heavy for a fewer number of robots. On the other hand, if there are too many robots assigned to one task, then resources will be wasted. In this example, if the robot-coalition size is less than four, then the coalition is useless. In order to address such scenarios effectively, our objective function (Definition $1$) requires the value of the coalition structure to be maximized, after minimizing the cost of forming it. Therefore, the region growing algorithm aims to optimize the value of the coalition structure found by the linear programming approach.

\begin{algorithm}[ht!]
\small{
\KwIn{$CS$: Current coalition structure (result of the linear programming solution); 
\\ \hskip0.35in $\mathcal{R}_{ua}$: unassigned robots.}
\KwOut{$CS'$: The final coalition structure.}
$T_{sort} \leftarrow$ Sort the tasks $T$ in descending order of the number of robots assigned to them.\\
\For{$t_i \in T_{sort}$}
	{
    	$c_i \leftarrow$ current coalition from $CS$ formed for task $t_i$\\
        \While{$|c_i| < O_i$}{
    		$rad \leftarrow length(cell)$\\
    		Grow a virtual ball of radius $rad$ around $t_i$\\
            \If{$r_j \in \mathcal{R}_{ua}$ AND $dist(r_j,t_i) \leq rad$}{
        		$c_i \leftarrow c_i \cup r_j$ {\small /*$CS$ is updated to $CS'$*/}\\
                $\mathcal{R}_{ua} \leftarrow \mathcal{R}_{ua} \setminus r_j$\\
        	}
    		$rad \leftarrow rad + length(cell)$\\
    	}
        \If{$|c_i| > O_i$}{
        	Remove the farthest robots $r_k \in c_i$ s.t. $|c_i| = O_i$ {\small /*$CS$ is updated to $CS'$*/}\\
            $\mathcal{R}_{ua} \leftarrow \mathcal{R}_{ua} \cup r_k$
        }
        
    }
    \caption{Region Growing algorithm for value optimization and assigning unassigned robots to tasks}
    \label{region growing}
    }
\end{algorithm}

The region growing algorithm is executed under one or both of the following conditions: first, the solution found in the previous stage is fractional; or second, for the coalition structure ($CS$) found by the linear programming solution, $Val(CS)\neq MAX\_VAL$. 
In the region growing process, a virtual ball (centered around a task) is iteratively grown. This ball decides which robots are ultimately clustered together for a particular task and which robots are removed from a cluster previously formed during the linear programming phase. This can happen if one coalition size was initially too large resulting from the previous solution. A ball is grown for each task. The region growing algorithm terminates when each robot is allocated to some task (i.e., assigned to a cluster).

Let $\mathcal{R}_{ua} \subseteq R$ denote the set of unassigned robots. One robot can be unassigned from any task because of one of the  following two reasons: first, a fractional solution has been found in the previous round for this particular robot; or second, the cluster previously formed is unnecessarily large for the task. Before the start of the region growing algorithm, the set $\mathcal{R}_{ua}$ is initialized with all robots unassigned to any cluster in the previous stage. 

In the region growing algorithm (shown as Algorithm 2), a virtual ball of a certain radius ($rad$) is grown for each task (with the task as its center) iteratively. Note that $rad$ is initialized to one $cell$ length. In other words, the ball will encompass all the robots which are one cell-away from the task $t_i$. In the next iteration, the radius is increased to two-cell length. If $t_i$'s virtual ball has already engulfed $O_i$ robots in it, then we stop growing the ball any further for this task. If there were more than $O_i$ robots assigned to this task, then those robots are declared unassigned now and added to the set $\mathcal{R}_{ua}$. Note that, the virtual ball of any task can engulf not only the already allocated robots to it in the linear programming phase, but also the robots which are part of the set $\mathcal{R}_{ua}$. 

\lemma{The worst-case time complexity of the region growing algorithm is $O(MN)$.}\\
{\em Proof: }The worst-case time complexity of the region growing algorithm is easily seen to be $O(MN)$ as follows. In step 1, the $M$ tasks are sorted in the descending order of the number of robots assigned. This will be of complexity $O(M\log M)$. The time complexity for the rest of the algorithm can be determined by observing that $|\mathcal{R}_{ua}|$, the number of unassigned robots must be reduced to zero. Of course, since $|\mathcal{R}_{ua}|\le N$, if $s$ iterations are required (by growing the region with increasing radii each time) for each coalition $c_i$, $i=1,2,\cdots,M$, the modification of $\mathcal{R}_{ua}$ and $c_i$ together will take at most $sMN=O(MN)$ time.
Thus, the time complexity of the region growing algorithm is $O(M\log M+MN)$. Since $N>M$, we conclude the complexity is $O(MN)$.

\lemma{Each task $t_i$ will get $O_i$ number of robots assigned to it.}\\
{\em Proof:} First, note that, $\sum\limits_{j=1}^{M} O_j = N$. This also means that, the total number of extra robots (based on $O$-value) assigned to some tasks is equal to the total number of less robots assigned to the rest of the tasks. Therefore, if any task $t_j$ is assigned more robots than $O_j$, then there is definitely one task $t_k$ for which $|c_k| < O_k$. When the tasks are sorted in descending order, the first task $t_1$ in the list $T_{sort}$ will have either exactly $O_1$ or more robots assigned to it. If $|c_1|=O_1$, then we move on to $t_2$. If $|c_1|>O_1$, then we detach the extra $(|c_1|-O_1)$ robots from it and put them into the set $\mathcal{R}_{ua}$. If $|c_j|<O_j$ for any task $t_j \in T_{sort}, j>1$, then robots from $\mathcal{R}_{ua}$ will be assigned to $t_j$. Thus, every task, $t_j$, will have exactly $O_j$ robots assigned to it, i.e., $|c_j| = O_j, \forall t_j \in T$.

\section{Experiments}
\subsection{Settings}
We have implemented our algorithms using the Java programming language, and ran tests on a desktop computer (Intel i7-7700 processor, 16GB RAM). We varied the number of robots ($N$) between $[10, 100]$ in steps of $10$ and selected the number of tasks ($M$) from the set $[2,4,6,8,10]$. Remembering that the maximum value of $S(N,M)$ could grow exponentially with increasing $M$, We have kept the number of tasks at a maximum level of 10 so that the number of possible partitions to consider does not become prohibitively large. Additionally, note that if the task count was greater than $50\%$ of the number of robots, that robot-task pair was not considered. The distinct $2$D locations of the robots and the tasks were randomly generated from $\mathcal{U}(\{[1,100],[1,100]\})$. 
Also, we considered all possible ways of assigning optimal numbers of robots for the tasks at hand. Thus, for each pair $(N,M)$ we considered for experimentation, we generated all possible $O_i$'s by partitioning $N$ into exactly $M$ parts using integer partitioning. For example, with $N=10$ and $M=2$, the set of $O_i$'s generated and tested are \{\{9,1\}, \{8,2\}, \{7,3\}, \{6,4\}, \{5,5\}\}. The results presented here represent the averages of the results obtained
over 10 runs with each of these settings.

\begin{figure}[ht!]
\begin{center}
\begin{tabular}{cc}
\hspace{-0.2in}\includegraphics[width=0.53\linewidth]{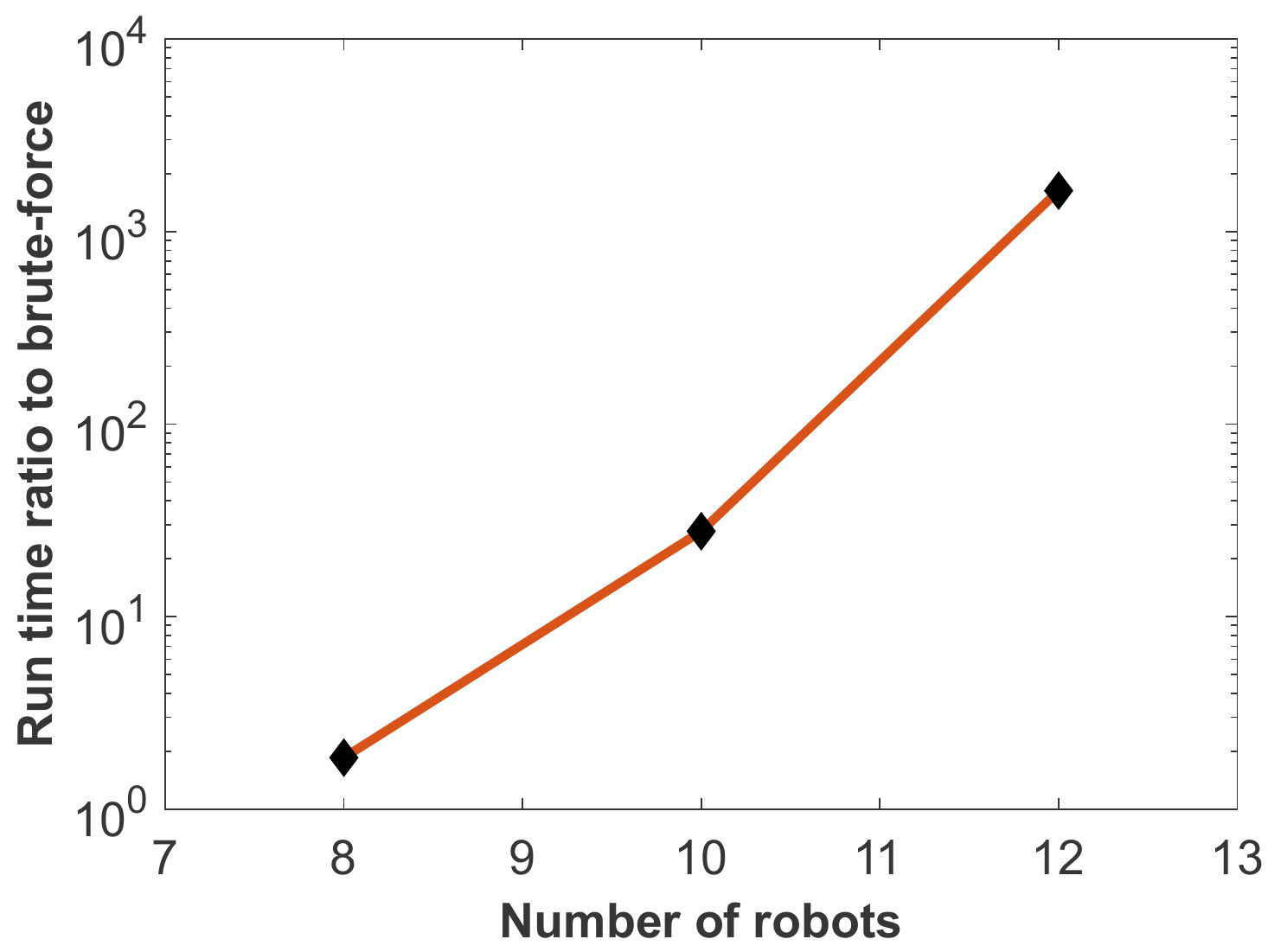}&
\hspace{-0.2in}\includegraphics[width=0.53\linewidth]{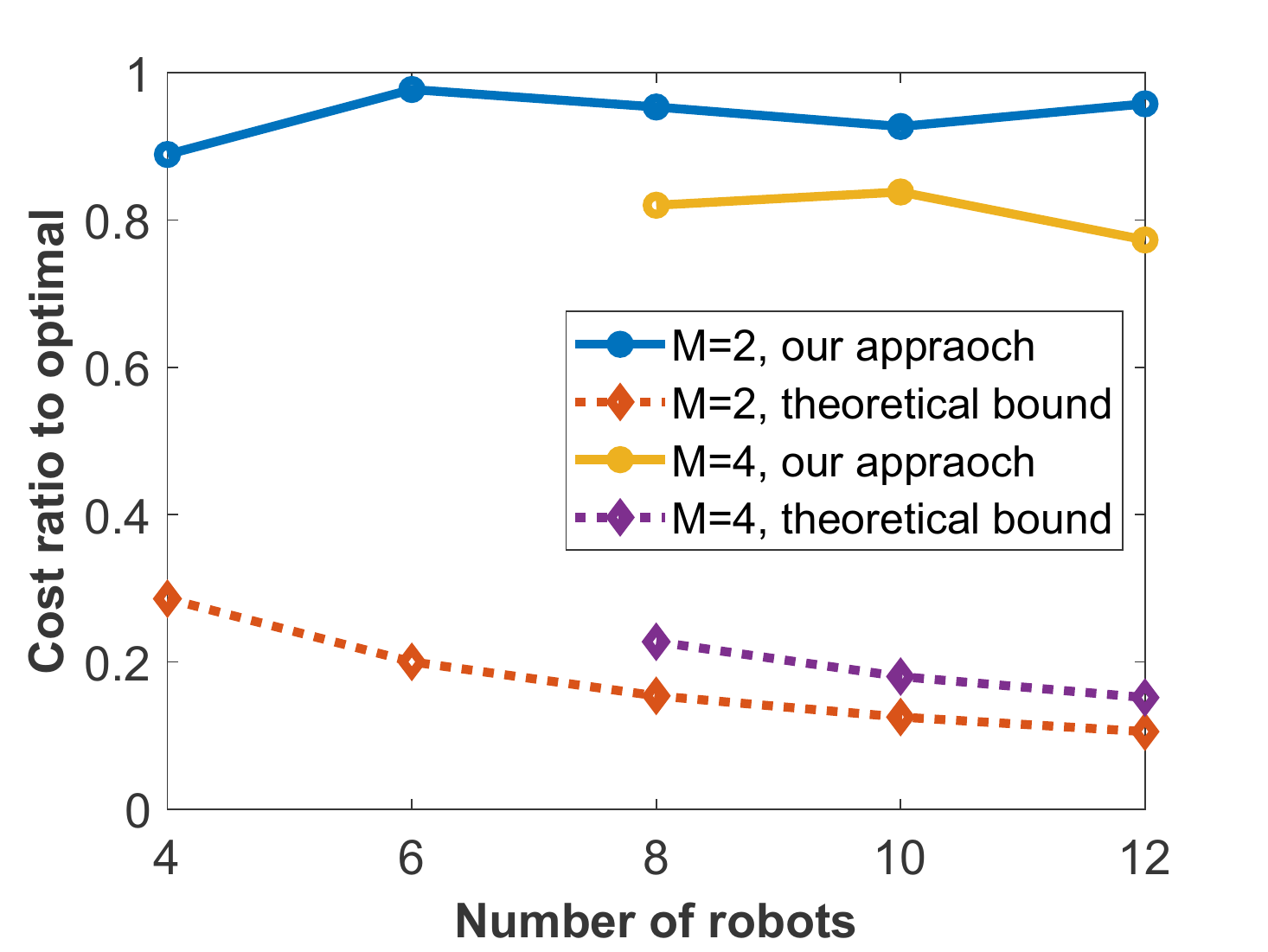}\\
(a)&(b)
\end{tabular}
\end{center}
\caption{(a) Runtime comparison on log-scale with a brute-force algorithm \cite{orlov2002efficient} ($4$ tasks); (b) Distance-based cost comparison with the optimal solution (higher is better -- $1$ being the best-case). Dotted lines indicate the theoretical bounds \cite{adams2011coalition} and the solid lines indicate the performances of our proposed approach.}
\label{brute-force_compare}
\end{figure}

\subsection{Results} In this section, we discuss our main findings from the experiments.\\
\textbf{Comparison with the optimal: }We have implemented a brute-force algorithm \cite{orlov2002efficient} that finds the optimal solution which can be compared with the solution produced by our proposed approach. As our proposed strategy always finds a solution with the maximum value, we found the coalition structure using the brute-force method which has the maximum value (using Eq.\ref{eqn_task_val}) and the minimum cost among all the maximum-valued coalition structures. We could test this algorithm for up to $12$ robots and $4$ tasks after which it became prohibitive on our test machine. Two metrics have been compared: 1) runtime and 2) total distance-cost among the robots and the tasks they are assigned to. The result is shown in Figure \ref{brute-force_compare}. As expected, this result (Fig. \ref{brute-force_compare}.(a)) shows that the brute-force algorithm takes considerably more time than our proposed approach (up to $1630$ times for $12$ robots and $4$ tasks). On the other hand, using our proposed approach, the robots need to travel almost similar amount of distances compared to the optimal solution. For example, with $6$ robots and $2$ tasks, the total distance traveled by the robots using the optimal solution was $296.49m.$; using our approach it was $303.58m.$; and this indicates a $97.66\%$ near-optimal result. Moreover, our proposed approach performs near-optimally in terms of finding the coalition structure with the lowest distance-cost measurement while keeping the value of the coalition structure optimal (Fig. \ref{brute-force_compare}.(b)). As the more distance traveled by the robots would result in higher battery expenditure, without loss of generality, we may claim that our proposed approach would eventually be able to bound the battery expenditure at a near-optimal. 

\begin{figure}[ht!]
\begin{center}
\begin{tabular}{cc}
\hspace{-0.2in}\includegraphics[width=0.53\linewidth]{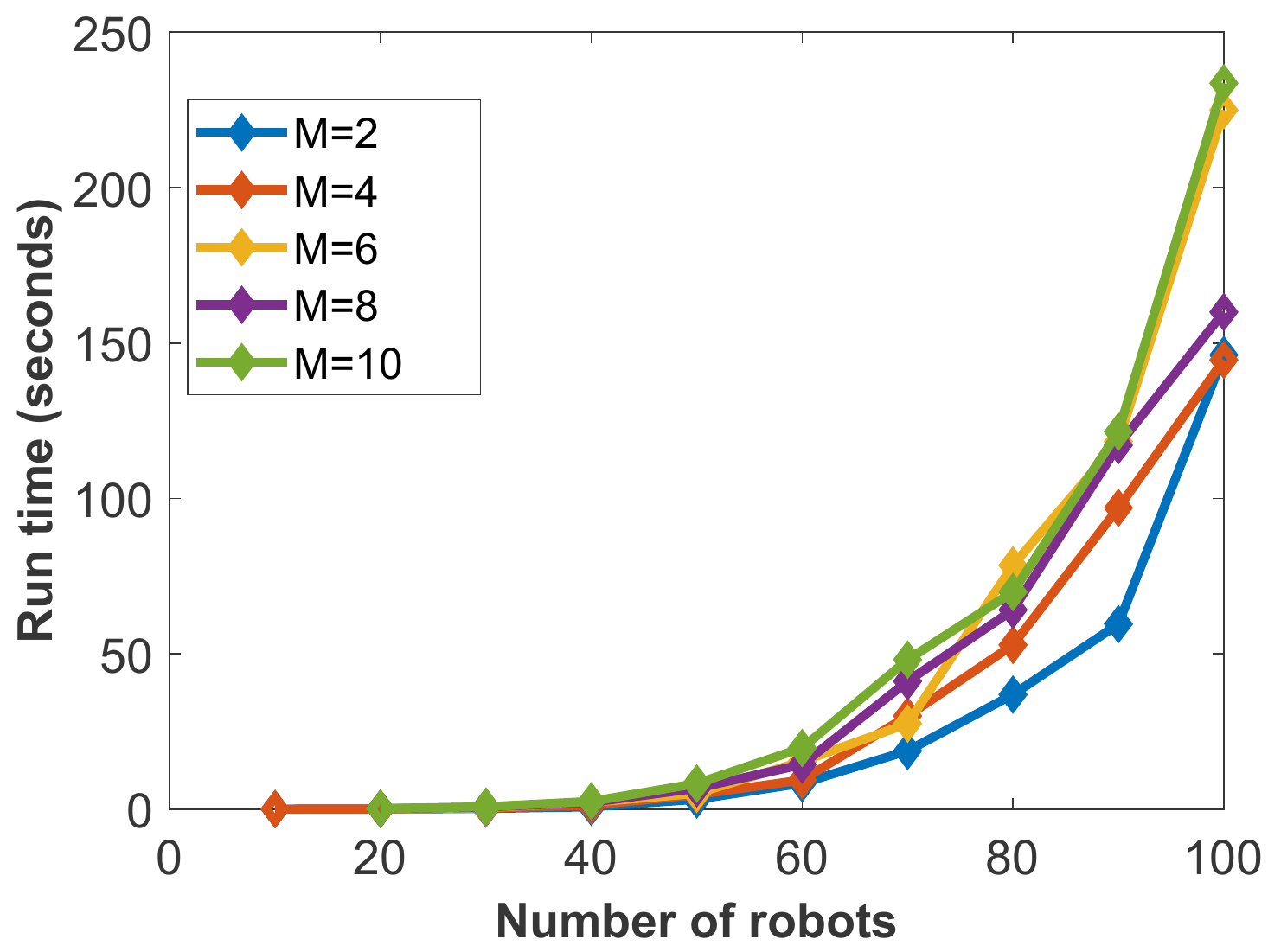}&
\hspace{-0.1in}\includegraphics[width=0.53\linewidth]{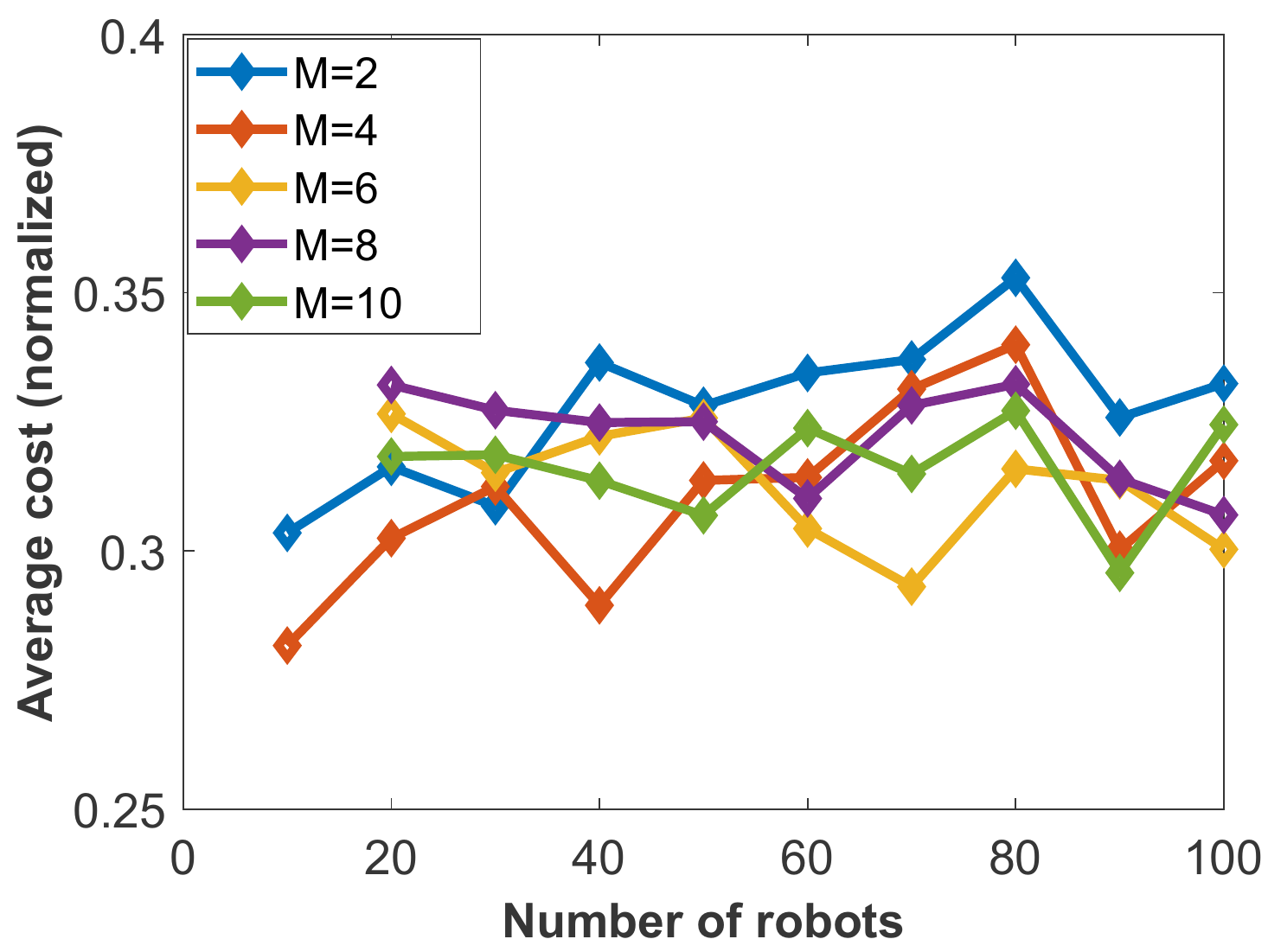}\\
(a)&(b)
\end{tabular}
\end{center}
\caption{(a) Runtime of our approach; (b) Normalized average distance-cost the robots.}
\label{RG_runtime}
\end{figure}

We also compare this distance-cost ratio to the optimal with a theoretical worst-case bound proved in \cite{Shehory98,adams2011coalition}. The plot of this theoretical bound $(\max_{i \in [1,m]} O_i +1)$ is shown in Fig. \ref{brute-force_compare}.(b). This figure shows that our method always finds a significantly better solution in terms of closeness to the optimal in each of the test cases. The maximum and the minimum difference of these two ratios are found to be $9.1$ times (for $12$ robots and $2$ tasks) and $3.61$ times (for $8$ robots and $4$ tasks).\newline
\textbf{Performance of our approach: }
Next, we show how the performance of our proposed approach scales with a large set of robots and tasks. First, we test how the runtime of our proposed algorithm scales for up to $100$ robots and $10$ tasks. As can be observed in Fig. \ref{RG_runtime}.(a), the maximum time taken by our approach is about $230$ secs. for $100$ robots being assigned to $10$ tasks. Note that, for this setting, the astronomical number of possible coalition structures is $2.75 \times 10^{93}$.

\begin{figure}[ht!]
\begin{center}
\begin{tabular}{cc}
\hspace{-0.2in}\includegraphics[width=0.53\linewidth]{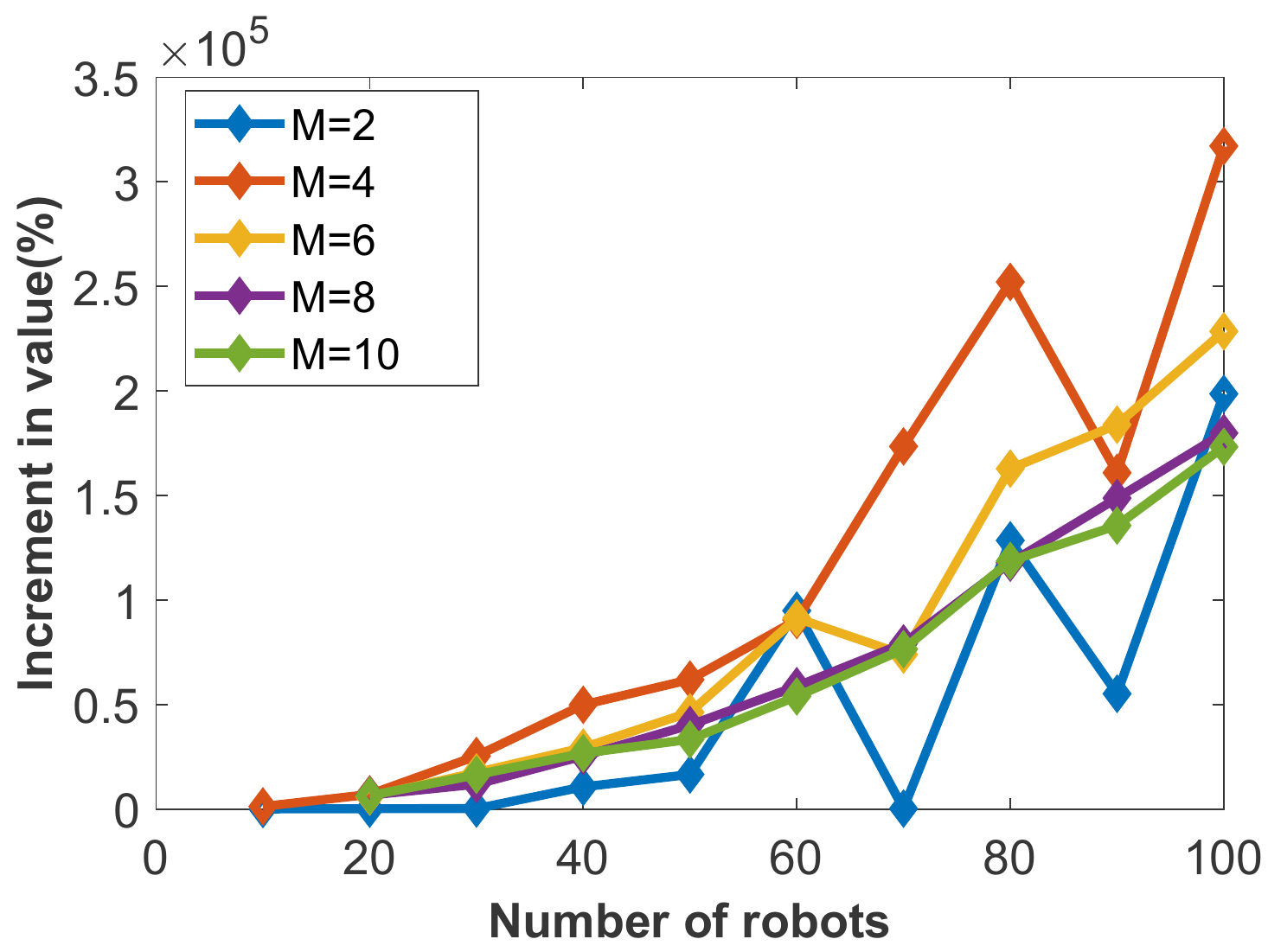}&
\hspace{-0.1in}\includegraphics[width=0.53\linewidth]{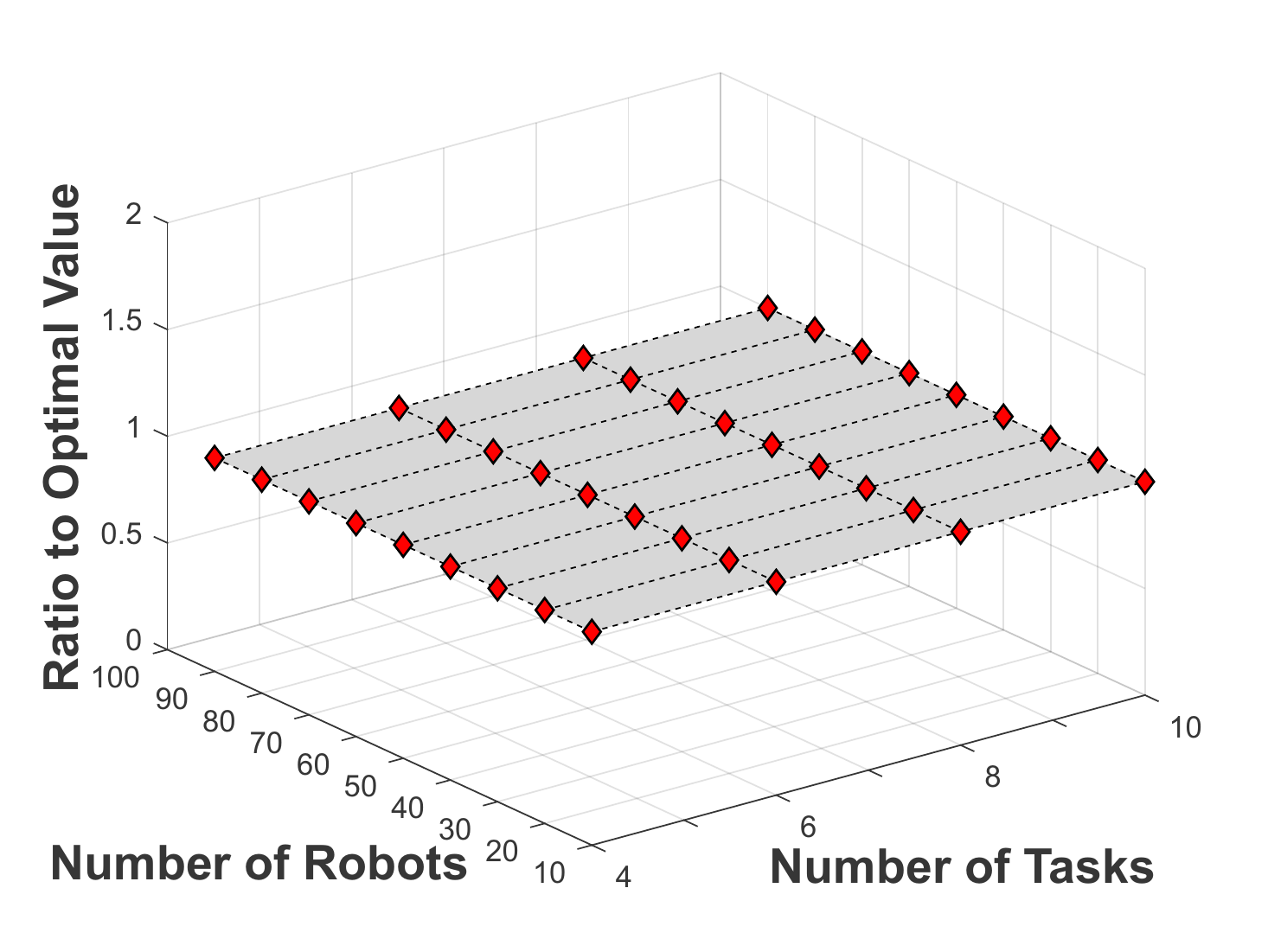}\\
(a)&(b)
\end{tabular}
\end{center}
\caption{(a) Increment in coalition structure value from the linear programming solution by using the region growing algorithm; (b) Comparison with the optimal value ($1$ being the best-case).}
\label{RG_value}
\end{figure}

In Fig. \ref{RG_runtime}.(b), we show how the normalized average distance ($\sum\limits_{r_i, t_j}cost_{dist}/N$) traveled by the robots for moving from their initial locations to their allocated task locations changes with different number of robots and tasks. 
In this figure, an almost-static  trend can be noticed while the maximum difference between any two cases being about $0.06$. This shows that the average distance traveled by the robots does not change significantly with varying $N$ and $M$. We are also interested to see how much gain we make in terms of the value of the a coalition structure by using the region growing algorithm. Remember that the linear programming component does not take the $O$-values into account while forming the best coalition structure. Therefore, we cannot guarantee that this coalition structure will have the value $MAX\_VAL$. It is evident from the result (Fig. \ref{RG_value}.(a)) that we always gain a significant amount of coalition structure value by using the region growing algorithm (up to $3.2 \times 10^5\%$). In Fig. \ref{RG_value}.(b), we see that the value of the coalition structure produced as a final output is always the optimal thus showing the importance of the region growing algorithm.

We have empirically demonstrated that although we minimize the cost and maximize the value of the coalition structures in two successive procedures, our approach could still form a coalition structure (i.e., a set of coalitions assigned to the tasks) which not only has the optimal value, the cost of it is also very close to the optimal. Also, this near-optimal result is produced in a negligible amount of time given the notoriously intractable nature of the problem.

\section{Conclusion and Future Work}
We have proposed a multi-robot coalition formation algorithm for task allocation inspired by the idea of correlated clustering. Our proposed approach first finds a coalition structure with the minimum cost using a linear programming-based graph partitioning formulation and next, using a region growing approach, it optimizes the value of this found coalition structure. We have empirically shown that our proposed approach can yield a near-optimal solution within an insignificant amount of time. This approach also performs significantly better compared to a previously proposed theoretical bound. In the future, we plan to make this approach distributed so that we can avoid the single point of failure. Also, we plan to implement this approach on a group of robots in a real-world setting. 

\bibliographystyle{abbrv}
\bibliography{references}
\end{document}